\documentclass[12pt]{article}
\usepackage{mathrsfs}
\usepackage{amsmath}
\usepackage{amssymb}

\usepackage{graphicx}
\usepackage{epstopdf}

\begin{document}

\title{Pion Model}
\author{B.G. Sidharth\\
B.M. Birla Science Centre,\\ Adarsh Nagar, Hyderabad - 500 063
(India)\footnote{email:iiamisbgs@yahoo.co.in}}
\date{}
\maketitle

\section{Introduction}
Yukawa introduced the concept of Pions as a possible mediator for
the strong forces, as then visualized. Later these were discovered.
They have been considered to be typical elementary particles because
of their role in strong interactions. Based on this Hayakawa
\cite{hayakawa1,hayakawa2} suggested a cosmological model that also
included weak interactions. So also the author put forward a
completely different cosmology in this journal \cite{ijmpa1998,mg8},
working with fluctuations and pions, thus correctly predicted the
accelerated expansion of the universe, when the standard Big Bang
model of that time stated
the exact opposite.\\
In later work \cite{bgs}the author has proposed a model, which
starts with the Quantum Vacuum pictured as a collection of zero
point oscillators or alternatively Planck scale oscillators
(Cf.ref.\cite{tduniv}). At this stage a Ginzburg-Landau phase
transition leads to the next stage or phase of the universe with
elementary
particles \cite{bgsepjc}. This can be seen as follows.\\
Let us consider an array of $N$ particles, spaced a distance $\Delta
x$ apart, which behave like oscillators that are connected by
springs. We then have \cite{bgsfpl152002,good,vandam,uof}
\begin{equation}
r  = \sqrt{N \Delta x^2}\label{4De1d}
\end{equation}
\begin{equation}
ka^2 \equiv k \Delta x^2 = \frac{1}{2}  k_B T\label{4De2d}
\end{equation}
where $k_B$ is the Boltzmann constant, $T$ the temperature, $r$ the
extent  and $k$ is the spring constant given by
\begin{equation}
\omega_0^2 = \frac{k}{m}\label{4De3d}
\end{equation}
\begin{equation}
\omega = \left(\frac{k}{m}a^2\right)^{\frac{1}{2}} \frac{1}{r} =
\omega_0 \frac{a}{r}\label{4De4d}
\end{equation}
We now identify the particles with \index{Planck}Planck
\index{mass}masses and set $\Delta x \equiv a = l_P$, the
\index{Planck}Planck length. It may be immediately observed that use
of (\ref{4De3d}) and (\ref{4De2d}) gives $k_B T \sim m_P c^2$, which
of course agrees with the temperature of a \index{black hole}black
hole of \index{Planck}Planck \index{mass}mass. Indeed, Rosen
\cite{rosen} had shown that a \index{Planck}Planck \index{mass}mass
particle at the \index{Planck scale}Planck scale can be considered
to be a \index{Universe}Universe in itself with a Schwarzchild
radius equalling the Planck length. Furthermore in the above
characterization a typical elementary particle like the
\index{pion}pion can be considered to be the result of $n \sim
10^{40}$ \index{Planck}Planck
\index{mass}masses.\\
Using this in (\ref{4De1d}), we get $r \sim l$, the \index{pion}pion
\index{Compton wavelength}Compton wavelength as required. Whence the
pion mass is given by
\begin{equation}
m = m_P/\sqrt{n}\label{ex}
\end{equation}
as indeed is the case. Further, in this latter case, using
(\ref{4De1d}) and the fact that $N = n \sim 10^{40}$, and
(\ref{4De2d}),i.e. $k_BT = kl^2/N$ and (\ref{4De3d}) and
(\ref{4De4d}), we get for a \index{pion}pion and (\ref{ex}),
$$k_ B T = \frac{m^3 c^4 l^2}{\hbar^2} = mc^2,$$
which of course is the well known formula for the Hagedorn
temperature for \index{elementary particles}elementary particles
like \index{pion}pions \cite{sivaramamj83}.\\
In this model we end up with the Hagedorn temperature \cite{tduniv}.
Now the importance of this is that in the Hagedorn theory Hadrons
condense at this temperature, and these Hadrons would now be pions.
In other words from the Quantum Vacuum through a phase transition we
are lead to the formation of pions \cite{nap2017}.
\section{A Mass Spectrum}
From these considerations various attempts have been made by several
authors for a simple or fundamental mass spectrum \cite{ijmpe2011}.
We note that all of non-leptonic matter is made up of quarks
\cite{le-81}
 They interact via the interquark or QCD force. In this picture, pions are bound-states of a quark and an anti
an anti-quark. The QCD potential is given by
\begin{equation}
U(r) = - \frac{\alpha}{r} + \beta r\label{4Ge1}
\end{equation}
where in units $\hbar = c = 1, \alpha \sim 1$. The first term in
(\ref{4Ge1}) represents the Coulombic part while the second term
represents the confining part of the potential. As is well known the
potential in (\ref{4Ge1}) explains two well known features viz.,
quark confinement and asymptotic freedom.\\
Let us consider the pion made up of two quarks along with a third
quark, one at the centre and two at the ends of an interval of the
order of the Compton wavelength, $r$. Then the central particle
experiences the force
\begin{equation}
\frac{\alpha}{(\frac{x}{2} + r)^2} - \frac{\alpha}{(\frac{x}{2} -
r)^2} \approx \frac{-2\alpha x}{r^3}\label{4Ge2}
\end{equation}
where $x$ is the small displacement from the mean position. Equation
(\ref{4Ge2}) gives rise to the Harmonic oscillator potential, and
the whole configuration resembles the
tri-atomic molecule.\\
Before proceeding we can make a quick check on (\ref{4Ge2}). We use
the fact that the frequency is given by
$$\omega = \left(\frac{\alpha^2}{m_\pi r^3}\right)^{\frac{1}{2}} = \frac{\alpha}
{(m_\pi r^3)^{\frac{1}{2}}}$$ whence the mass of the pion $m_\pi$ is
given by
\begin{equation}
(h \omega \equiv ) \omega = m_\pi\label{4Ge3}
\end{equation}
Remembering that $r = 1/m_\pi$, use of (\ref{4Ge3}) gives $\alpha =
1$, which of course is
correct.\\
To proceed, the energy levels of the Harmonic oscillator are now
given by,
$$\left(n + \frac{1}{2}\right) m_\pi ,$$
If there are $N$ such oscillators, then over the various modes the
energy of the particle is given by
$$E = \sum^{3N}_{r=1} \left(n_r + \frac{1}{2}\right) \hbar \omega =
l \left(n + \frac{1}{2}\right) \hbar \omega$$ $m$ and $n$ being
positive integers. The mass of the particle $P$ is now given by
\cite{bgs-hadronic-04,uof,tduniv}
\begin{equation}
m_P = l \left(n + \frac{1}{2}\right) m_\pi\label{4Ge4}
\end{equation}
It is remarkable that the above simple formula reproduces for the
different integral values of $l$ and $n$ reproduces the masses of
all the known Bosons with an error of less than 1\% in most of the
cases and exceptionally less than about 2\%.\\
The above theory suggests that these particles are composed of (or
can decompose into) pions. An immediate example is the $K_0$ Meson
which can break into
two or even three pions.\\
More generally in a collision of the above particle, at high
energies we can expect the following

\begin{equation}\label{eq-5} A + B \rightarrow C+D + \Delta \end{equation}
where $A,\,B,\,C, $ are particles from the above table, $D$ are
pions if any and $\Delta$ is the energy (or conversely).
\section{Remarks}
\begin{enumerate}
  \item Interestingly six decades ago Nambu had given  the empirical formula $n \, \mbox{or} \, (n+\frac{1}{2}) \times (\mbox{the mass of the pion})$ for
  about half a dozen particles,
  remembering that at that time such a small number of particles were known \cite{na-52}. However this formula was nothing more than a curiosity because
  it was completely ad hoc  and devoid of any dynamics.

  \item Just as the nucleus can be split into sub-constituents we could conceive of an elementary particle being split into its constituent quarks
  with the release of energy. This however is forbidden by the quark confining force in the QCD potential. The above considerations on the other hand suggest that
  the particles could be split or fused into other particles and pions with the release of energy.

  \item We quickly observe that the above mass spectrum formula (\ref{4Ge4}) works remarkably well for all known non-leptonic Fermions
  too \cite{tduniv}. This appears surprising because it must be borne in mind that this derivation is insensitive to quantum numbers such as spin and so on.
\end{enumerate}

On the other hand, pions being Hadrons, it is expected that they
contribute to the masses of the other Hadrons. In any case it is
expected that reactions (\ref{eq-5}) hold good for Fermions as well,
provided that there is a balance of other quantum numbers. Indeed,
as is well known such pion production has been
observed at the RHIC, over the years in proton-proton and other collisions \cite{kanazawa-00,levaieur-08,levaiar-08b}.\\

In any case the very fact that all non leptonic particle masses are
covered (with error limits), by a single formula in terms of the
pion mass is in itself very suggestive and interesting. In other
words (\ref{4Ge4}) holds good for all known elementary particles.
Further (\ref{eq-5}) maybe important in the context of very high
energy collisions taking place at 14 TeV expected at LHC in 2013.

\vfill \LARGE

\noindent {\large{\bf APPENDIX}}

\begin{table}[tbp]\label{table1}
\caption{Mesons}%
\begin{tabular}{|c|c|c|c|}
\hline Particle and mass & Mass From Formula & Error \% &
$(\textit{l},n)$
\\ \hline
$\pi^{\pm} (139.57018) \pm 0.00035$ & $137$ & $-1.43885$ & $(2,0)$ \\
$\pi^0 (134.9766) \pm 0.0006$ & $137$ & $1.481481$ & $(2,0)$ \\
$K^{\pm} 4064 \pm 0.016$ & $496$ & $1.9$ & $(1,3)$ \\
$\eta (547.51) \pm 0.18$ & $548$ & $0.182815$ & $(8,0)$ \\
$f_0(600) (400-1200)$ & $616.5$ & $(2.75)0$ & $(1,4)$ \\
$\rho (775.5) \pm 0.4$ & $753.5$ & $-2.14286$ & $(1.5)$ \\
$\omega (782) \pm 0.12$ & $753.5$ & $-3.6445$ & $(1,5)$ \\
$\eta^{\prime}(958) \pm 0.14$ & $959$ & $0.104384$ & $(2,3)$ \\
$f_0(980) \pm 10$ & $959$ & $-2.14286$ & $(2,3)$ \\
$a_0(980) \pm 1.2$ & $959$ & $-2.14286$ & $(2,3)$ \\
$\phi (1020) \pm 0.020$ & $1027.5$ & $0.735294$ & $(1,7)$ \\
$h_1 (1170) \pm 20$ & $1164.5$ & $(-0.47009)0$ & $(1,8)$ \\
$b_1 (1235) \pm 3.2$ & $1233$ & $(-0.16194)0$ & $(2,4)$ \\
$a_1 (1260) 1230\pm 40$ & $1233$ & $(-2.14286)0$ & $(2,4)$ \\
$f_2 (1270) 1275.4\pm 1.1$ & $1233$ & $-2.91339$ & $(2,4)$ \\
$f_1 (1285) 1281.8\pm 0.6$ & $1301.5$ & $1.284047$ & $(1,9)$ \\
$\eta (1295)1294\pm 4$ & $1301.5$ & $0.501931$ & $(1,9)$ \\
$\pi (1300) \pm 100$ & $1301.5$ & $0.115385$ & $(1,9)$ \\
$a_2 (1320)1318.3\pm 0.6$ & $1301.5$ & $-1.40152$ & $(1,9)$ \\
$f_0 (1370) (1200-1500)$ & $1370$ & $0$ & $(4,2)$ \\
$h_1 (1380)$ & $1370$ & $0.72464$ & $(4,2)$ \\
$\pi_1 (1400) 1376\pm 17$ & $1370$ & $-2.14286$ & $(4,2)$ \\
$f_1 (1420) 1426.3\pm 0.9$ & $1438.5$ & $1.302817$ & $(1,10)$ \\
$\omega (1420) 1400-1450$ & $1438.5$ & $(1.302817)0$ & $(1,10)$ \\
$f_2 (1430)$ & $1438.5$ & $0.594406$ & $(1,10)$ \\
$\eta (1440) (1400-1470)$ & $1438.5$ & $-0.10417$ & $(1,10)$ \\
$a_0 (1450) 1474\pm 19$ & $1438.5$ & $-0.7931$ & $(1,10)$ \\
$\rho (1450) 1459\pm 11$ & $1438.5$ & $-0.7931$ & $(1,10)$ \\
$f_0 (1500) 1507\pm 5$ & $1507$ & $(0.466667)0$ & $(2,5)$ \\
$f_1 (1510)$ & $1507$ & $-0.19868$ & $(2,5)$ \\
$f^{\prime}_2 (1525) \pm 5$ & $1507$ & $-1.18033$ & $(2,5)$ \\
$f_2 (1565)$ & $1575.5$ & $0.670927$ & $(1,11)$ \\ \hline
\end{tabular}%
\end{table}

\begin{table}[tbp]
\begin{tabular}{|c|c|c|c|}
\hline Particle and mass & Mass From Formula & Error \% &
$(\textit{l},n)$
\\ \hline
$h_1 (1595)$ & $1575.5$ & $-1.22257$ & $(1,11)$ \\
$\pi_1 (1600)1653\pm 8$ & $1575.5$ & $-1.53125$ & $(1,11)$ \\
$\chi (1600)$ & $1575.5$ & $-1.53125$ & $(1,11)$ \\
$a_1 (1640)$ & $1644$ & $0.243902$ & $(8,1)$ \\
$f_2 (1640)$ & $1644$ & $0.243902$ & $(8,1)$ \\
$\eta_2 (1645)1617 \pm5$ & $1644$ & $(0.06079)0$ & $(8,1)$ \\
$\omega (1670)$ & $1644$ & $(1.55688)0$ & $(8,1)$ \\
$\omega_3 (1670) 1667\pm 4$ & $1644$ & $-1.55689$ & $(8,1)$ \\
$\pi_2 (1670) 1672.4\pm 3.2$ & $1644$ & $-1.55689$ & $(8,1)$ \\
$\phi (1680) \pm 20$ & $1712.5$ & $1.934524$ & $(1,12)$ \\
$\rho_3 (1690) 1688.8\pm 2.1$ & $1712.5$ & $1.331361$ & $(1,12)$ \\
$\rho (1700)1720\pm 20$ & $1712.5$ & $(0.735294)0$ & $(1,12)$ \\
$a_2 (1700)$ & $1712.5$ & $0.735294$ & $(1,12)$ \\
$f_0(1710) 1718\pm 6$ & $1712.5$ & $(0.146199)0$ & $(1,12)$ \\
$\eta (1760)$ & $1781$ & $1.193182$ & $(2,6)$ \\
$\pi (1800)1812\pm 14$ & $1781$ & $-1.05556$ & $(2,6)$ \\
$f_2(1810)$ & $1781$ & $-1.60221$ & $(2,6)$ \\
$\phi_3 (1850)1854\pm 7$ & $1849.5$ & $(-0.02703)0$ & $(1,13)$ \\
$\eta_2 (1870)$ & $1849.5$ & $-1.09626$ & $(1,13)$ \\
$\rho (1900)$ & $1918$ & $0.947368$ & $(4,3)$ \\
$f_2 (1910)$ & $1918$ & $0.418848$ & $(4,3)$ \\
$f_2 (1950) 1944\pm 12$ & $1918$ & $-1.64103$ & $(4,3)$ \\
$\rho_3 (1990)$ & $1986.5$ & $-0.17588$ & $(1,14)$ \\
$X (2000)$ & $1986.5$ & $-0.675$ & $(1,14)$ \\
$f_2 (2010) (+60/-80)$ & $1986.5$ & $(-1.16915)0$ & $(1,14)$ \\
$f_0 (2020)$ & $1986.5$ & $1.65842$ & $(1,14)$ \\
$a_4 (2040)2001\pm 10$ & $2055$ & $0.735294$ & $(2,7)$ \\
$f_4 (2050) 2025\pm 10$ & $2055$ & $0.243902$ & $(2,7)$ \\
$\pi_2 (2100)$ & $2123.5$ & $1.119048$ & $(1,15)$ \\
$f_0 (2100)$ & $2123.5$ & $1.119048$ & $(1,15)$ \\
$f_2 (2150)$ & $2123.5$ & $-1.23256$ & $(1,15)$ \\ \hline
\end{tabular}%
\end{table}

\begin{table}[tbp]
\begin{tabular}{|c|c|c|c|}
\hline Particle and mass & Mass From Formula & Error \% &
$(\textit{l},n)$
\\ \hline
$\rho_2 (2150)$ & $2123.5$ & $-1.23256$ & $(1,15)$ \\
$f_0 (2200)$ & $2260.5$ & $2.75$ & $(1,16)$ \\
$f_J (2220)$ & $2260.5$ & $1.824324$ & $(1,16)$ \\
$\eta (2225)$ & $2360$ & $1.595506$ & $(1,16)$ \\
$\rho_3 (2250)$ & $2260$ & $0.466667$ & $(1,16)$ \\
$f_2 (2300) 2297\pm 28$ & $2329$ & $1.26087$ & $(2,8)$ \\
$f_4 (2300)$ & $2329$ & $1.26087$ & $(2,8)$ \\
$D_s (2317)$ & $2329$ & $0.5$ & $(2,8)$ \\
$f_0 (2330)$ & $2329$ & $-0.04292$ & $(2,8)$ \\
$f_2 (2340)2339\pm 60$ & $2329$ & $-0.47009$ & $(2,8)$ \\
$\rho_5 (2350)$ & $2329$ & $-0.89362$ & $(2,8)$ \\
$a_6 (2450)$ & $2466$ & $-0.89362$ & $(4,4)$ \\
$f_6 (2510)$ & $2534.5$ & $0.976096$ & $(1,18)$ \\
$K^* (892) \pm 0.26$ & $890.5$ & $-0.16816$ & $(1,6)$ \\
$K_1(1270) 1272\pm 7$ & $1233$ & $2.91338$ & $(2,4)$ \\
$K_1(1400) 1402\pm 7$ & $1370$ & $-2.14286$ & $(4,2)$ \\
$K^*(1410) 1414\pm 15$ & $1438.5$ & $2.021277$ & $(1,10)$ \\
$K^*_0(1430) 1414\pm 6$ & $1438.5$ & $0.594406$ & $(1,10)$ \\
$K^*_2(1430) 1425.6\pm 1.5$ & $1438.5$ & $0.594406$ & $(1,10)$ \\
$K (1460)$ & $1438.5$ & $-1.4726$ & $(1,10)$ \\
$Pentaquark(1.5GeV)$ & $1.5$ & $0$ & $(2,5)$ \\
$K_2(1580)$ & $1575.5$ & $-0.28481$ & $(1,11)$ \\
$K (1630)$ & $1644$ & $0.858896$ & $(8,1)$ \\
$K_1 (1650)$ & $1644$ & $-0.36364$ & $(8,1)$ \\
$K^* (1680) 1717\pm 27$ & $1712.5$ & $(1.934524)0$ & $(1,12)$ \\
$K_2 (1770) 1773\pm 8$ & $1781$ & $(0.621469)0$ & $(2,6)$ \\ \hline
\end{tabular}%
\end{table}

\begin{table}[tbp]
\begin{tabular}{|c|c|c|c|}
\hline Particle and mass & Mass From Formula & Error \% &
$(\textit{l},n)$
\\ \hline
$K^*_3 (1780) \pm 7$ & $1781$ & $(0.05618)0$ & $(2,6)$ \\
$K_2 (1820) \pm 13$ & $1849.5$ & $1.620879$ & $(1,13)$ \\
$K (1830)$ & $1849.5$ & $1.065574$ & $(1,13)$ \\
$K^*_0 (1950)$ & $1918$ & $-1.64103$ & $(4,2)$ \\
$K^*_2 (1980)$ & $1986.5$ & $0.328283$ & $(1,14)$ \\
$K^*_4 (2045) \pm 9$ & $2055$ & $(0.488998)0$ & $(2,7)$ \\
$K_2 (2250)$ & $2260.5$ & $0.466667$ & $(1,16)$ \\
$K_3 (2320)$ & $2329$ & $0.387931$ & $(2,8)$ \\
$K^*_5 (2380)$ & $2397.5$ & $0.735294$ & $(1,17)$ \\
$K_4 (2500)$ & $2466$ & $-1.36$ & $(4,4)$ \\
$K (3100)$ & $3082.5$ & $-0.56452$ & $(1,22)$ \\
$D^\pm (1869.3)$ & $1849.5$ & $-1.05922$ & $(1,13)$ \\
$D^\pm_0 (1968.5) \pm 0.6$ & $1986.5$ & $0.914402$ & $(1,14)$ \\
$D^*_0 (2007)2006.7\pm 0.4$ & $1986.5$ & $-1.02143$ & $(1,14)$ \\
$D^*_\pm (2010) \pm 0.4$ & $1986.5$ & $-1.16915$ & $(1,14)$ \\
$D_S (2317) 2317.3 \pm 0.6$ & $2329$ & $0.51791$ & $(2,8)$ \\
$D_1 (2420) 2422.3\pm 1.3$ & $2397.5$ & $-0.92975$ & $(1,17)$ \\
$D_1^\pm (2420)$ & $2397.5$ & $-0.97067$ & $(1,17)$ \\
$D^*_2 (2460)2461.1\pm 1.6$ & $2466$ & $0.243902$ & $(4,4)$ \\
$D^*_\pm (2460) 2459 \pm 4$ & $2466$ & $0.243902$ & $(4,4)$ \\
$D^\pm_{S1} (2536)2535.35 \pm 0.34$ & $2534.5$ & $-0.07885$ & $(1,18)$ \\
$D_{SJ} (2573) \pm 1.5$ & $2534.5$ & $-1.49631$ & $(1,18)$ \\
$B^{\pm} (5278) 2579\pm 0.5$ & $5274.5$ & $-0.08524$ & $(1,38)$ \\
$B^0 (5279.4) \pm 0.5$ & $5274.5$ & $-0.09281$ & $(1,38)$ \\
$B_j(5732)$ & $5754$ & $-0.47009$ & $(4,10)$ \\
$B^0_S (5369.6) 5367.5\pm 1.8$ & $5343$ & $-0.49538$ & $(2,19)$ \\
$B^*_{SJ} (5850)$ & $5822.5$ & $-0.47009$ & $(1,42)$ \\
$B^\pm_c (6400) 6286\pm 5$ & $6370.5$ & $0.4609$ & $(3,15)$ \\
$\eta c (1S) (2979) 2980.4 \pm 1.2$ & $2945.5$ & $-1.12454$ & $(1,21)$ \\
$J/\psi (1S) (3096)3096.916 \pm 0.011$ & $3082.5$ & $-0.46402$ & $(1,22)$ \\
$\chi c_0 (1P) (3415.1) 3414.76\pm 0.35$ & $3425$ & $0.289889$ & $(2,12)$ \\
$\chi c_1 (1P) (3510.5) \pm 0.07$ & $3493.5$ & $-0.48426$ & $(1,25)$ \\
$\chi c_2 (1P) (3556) 3556.20\pm 0.09$ & $3562$ & $0.168729$ & $(4,6)$ \\
\hline
\end{tabular}%
\end{table}

\begin{table}[tbp]
\begin{tabular}{|c|c|c|c|}
\hline Particle and mass & Mass From Formula & Error \% &
$(\textit{l},n)$
\\ \hline
$\psi (2S) (3685.9)3686.093 \pm 0.034$ & $3699$ & $0.355408$ & $(2,13)$ \\
$\psi (3770)3771.1 \pm 2.4$ & $3767.5$ & $(-0.06631)0$ & $(1,27)$ \\
$\psi (3836)$ & $3836$ & $0$ & $(8,3)$ \\
$\chi (3872)3871.2\pm 0.5$ & $3876$ & $0.13$ & $(3,9)$ \\
$\chi_{c2} (28) 3929\pm 5$ & $3944$ & $0.38$ & $(2,14)$ \\
$\psi (4040) 4039\pm 1$ & $4041.5$ & $(0.037129)0$ & $(1,29)$ \\
$\psi (4160) 4153\pm 3$ & $4178.5$ & $(0.444712)0$ & $(1,30)$ \\
$\psi (4415) 4421\pm 4$ & $4452.5$ & $0.84937$ & $(1,32)$ \\
$\gamma (1S) (9460.3) \pm 0.26$ & $9453$ & $-0.07716$ & $(2,34)$ \\
$\chi b_0 (1P) (9859.9)9859.4\pm 0.42\pm 0.31$ & $9864$ & $0.041583$ & $%
(16,4)$ \\
$\chi b_1 (1P) (9892.7) \pm 0.6$ & $9864$ & $-0.29011$ & $(16,4)$ \\
$\chi b_2 (1P) (9912.6) \pm 0.5$ & $9864$ & $-0.49029$ & $(16,4)$ \\
$\gamma (2S) (10023) \pm 0.00031$ & $10001$ & $0.21949$ & $(2,36)$ \\
$\chi b_0 (2P) (10232) \pm 0.0006$ & $10275$ & $0.42026$ & $(2,37)$ \\
$\chi b_1 (2P) (10255) \pm 0.0005$ & $10275$ & $0.1945027$ & $(2,37)$ \\
$\chi b_2 (2P) (10268) \pm 0.0004$ & $10275$ & $0.068173$ & $(2,37)$ \\
$\gamma (3S) (10355) \pm 0.0005$ & $10343.5$ & $0.11105$ & $(1,75)$ \\
$\gamma (4S) (10580) 10579.4\pm 1.2$ & $10549$ & $-0.29301$ & $(2,38)$ \\
$\gamma (10860) 10865\pm 8$ & $10891.5$ & $0.290055$ & $(3,26)$ \\
$\gamma (11020) 11019\pm 8$ & $11028.5$ & $0.077132$ & $(1,80)$ \\
\hline
\end{tabular}%
\end{table} \eject

\end{document}